\documentstyle[prl,aps,epsf]{revtex}
\draft
\begin{document}
\twocolumn[\hsize\textwidth\columnwidth\hsize\csname @twocolumnfalse\endcsname

\title{Rectification and Phase Locking For 
Particles on Two Dimensional Periodic Substrates 
}
\author{C. Reichhardt, C.J. Olson and M.B.~Hastings} 
\address{ 
Center for Nonlinear Studies, Theoretical, and  Applied 
Physics Division, 
Los Alamos National Laboratory, Los Alamos, NM 87545}

\date{\today}
\maketitle
\begin{abstract}
We show that a novel rectification phenomena is possible for 
overdamped particles
interacting with a 2D periodic substrate and driven with  
a longitudinal DC drive and a circular AC drive. 
As a function of DC amplitude, the longitudinal velocity 
increases in a series of quantized steps with  
transverse 
rectification occurring  near these transitions. 
We present 
a simple model that captures the quantization 
and rectification behaviors. 
\end{abstract}
\vspace{-0.1in}
\vspace{-0.3in}

\vskip2pc]
\narrowtext

Particles transported through
periodic potentials 
under an AC drive exhibit a remarkable variety of
nonlinear behavior. 
One of the most 
intensely studied phenomena found in these systems
is the phase-locking that occurs
when the external AC frequency matches the frequency of 
motion over the periodic potential, such as the well-known    
Shapiro steps observed
in the V(I) curves of Josephson-junction arrays
\cite{Shapiro}. 
Phase locking steps in transport curves are also seen in other
systems such as charge density waves \cite{Thorne} and 
vortices in superconductors
with periodic substrates \cite{Martinoli,Look,Zimanyi}.  
If the
potential through which the particle moves is asymmetric, a
rectification or 
ratchet effect can occur \cite{Rachet,Atom,Det,FluxRatchet,Hanggi}
in both thermal \cite{Rachet,Atom} and 
deterministic ratchets \cite{Det}. 
The ratchet effect has been studied in the
context of biological motors \cite{Rachet} and 
can be utilized for practical applications such as particle segregation 
\cite{Rachet,Det}, atom transport in optical lattices 
\cite{Atom}, 
and fluxon removal in superconducting
devices \cite{FluxRatchet,Hanggi}. 
In deterministic ratchets, as a 
function of AC amplitude, a complex series of phases appear in which
rectification in either direction
and current reversals can occur 
\cite{Det}. 

In this work we study a novel form of rectification and
phase locking for 
an overdamped particle without thermal effects
moving in a 2D periodic substrate 
under an applied longitudinal
DC drive (${\bf f}_{DC}$) and two AC drives: 
${\bf f}_{AC}^{x}$
in the longitudinal direction 90 degrees out of phase from
${\bf f}_{AC}^{y}$
in the transverse direction.
For AC amplitude $A$ large enough that
the stationary (${\bf f}_{DC}=0$) particle orbit encircles at least one 
substrate potential maximum, we find
that the longitudinal velocity increases in a series of  
steps as ${\bf f}_{DC}$ increases,
and that near the transition between two steps 
a rectification in the 
{\it transverse} direction occurs. 
The steps correspond to drives
at which the 
particle motion forms orbits commensurate with
the substrate period.
The transverse rectification occurs when
the DC drive imposes an asymmetry on the particle orbit, 
interfering with the commensuration and allowing particles to ``leak''
through the substrate in the transverse direction.
The rectification is predominantly in one direction; however, 
reversals of the rectification also occur.  
We capture the qualitative features of the steps and rectification
with a simple model, and show that our results are generic. 
The behavior we describe can be observed in vortices in 
superconductors with periodic pinning, 
as well as overdamped charged  
particles such as colloids interacting with periodic substrates.   
In addition, 
the rectification could be
useful for particle species segregation, such as in
polydisperse colloidal systems. 

As a model for vortices in superconductors or colloids in solution,
we consider an overdamped particle moving in 2D interacting with
a repulsive periodic substrate according to the equation of motion: 
$ {\bf f}_{i} =  
{\bf f}_{s} + {\bf f}_{DC} + {\bf f}_{AC} = 
\eta{\bf v}_{i}$, with 
$\eta=1$. 
The force from the substrate, a square array of side $a$, 
is ${\bf f}_{s} = -\nabla U(r)$; the form of $U(r)$ is
discussed below.
We consider a system of size $8a \times 8a$.
The DC drive ${\bf f}_{DC}$ is applied
along the symmetry axis of the  
pinning array, in the $x$ or longitudinal direction. 
The AC drive
is ${\bf f}_{AC} = A\sin(\omega_{A}t){\hat {\bf x}} + 
B\cos(\omega_{B}t){\hat {\bf y}}$. Note that there is {\it no} 
DC driving component in the $y$ or transverse direction. 
We fix
$w_{A}/w_{B} = 1.0$ and $A = B$, and 
examine both the longitudinal 
time averaged particle velocity $<V_{x}>$
and the transverse velocity $<V_{y}>$.   
${\bf f}_{DC}$ is increased from 0 to $1.0$
in increments of $0.0001$,
with $3 \times 10^5$ time steps spent at each drive
to ensure a steady state. 

To model specific systems, we consider
periodic substrate potentials created by pinned particles, such
as vortices in a periodic array of holes \cite{Moshchalkov} or
magnetic dots \cite{Schuller}. 
Once all the holes are filled with a vortex, 
any remaining vortices sit in 
the interstitial regions between holes.
An unpinned vortex experiences a
smooth periodic 
substrate created by the interactions with the pinned vortices. 
Phase locking of AC and DC driven interstitial vortices can occur, and 
has been observed 
in experiments \cite{Look} and simulations \cite{Zimanyi}. 
For ${\bf f}_{AC}=0$, 
there is a finite DC depinning threshold for the interstitial vortices.
For ${\bf f}_{DC}=0$, two crossed AC drives with small amplitudes $A$ 
cause the vortex to move in a circle in the interstitial regions. 
At increasing $A$, there are  stable vortex orbits which encircle
one pinned vortex, then four, nine, and so on. 
We concentrate
on $A$ large enough 
to generate orbits encircling one or more pinned vortices.
For the pinned particles, we use the potential 
for vortices in a thin film
superconductor, $U(r) = -\ln (r)$, 
and employ a 

\begin{figure}
\center{
\epsfxsize=3.5in
\epsfbox{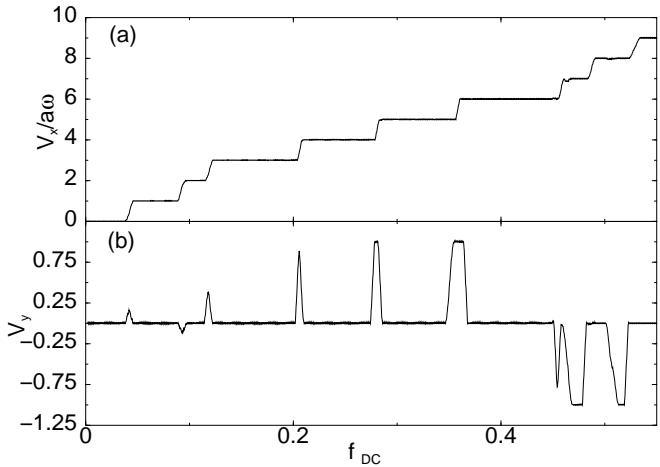}}
\caption{
(a) Longitudinal velocity $V_{x}$ vs driving force ${\bf f}_{DC}$ which is 
applied in the $x$ direction. 
(b) Corresponding transverse velocity $V_{y}$ vs ${\bf f}_{DC}$. 
}
\end{figure}

\hspace{-13pt}
summation technique \cite{Gronbech} 
for the long range 
interaction. 
We have also considered potentials 
for
unscreened or screened charges with interaction
of $U(r) = 1/r$ and $U(r)=e^{-\kappa r}/r$, respectively. 

In Fig.~1(a) we show the longitudinal velocity 
$V_{x}/(a \omega)$ versus ${\bf f}_{DC}$
for a particle in a system with $U(r)=-\ln(r)$,
and $A = B = 0.36$. For these parameters, at ${\bf f}_{DC}=0$ the particle
encircles one pin for $0.29 < A < 0.4$.    
As shown, $V_{x}$ increases in quantized steps with step heights 
of $\Delta V_{x} = a\omega$.  We label a step $n$ according to the
value of $V_{x}$ on the step, $V_{x}=na\omega$.
The widths of the integer 
steps varies with $A$ and $\omega$ in 
an oscillatory manner. 
There are also some fractional 
steps with heights $\Delta V_{x} = (p/q)a\omega $ where $p$
and $q$ are integers such that $p/q < 1.0$. 

In Fig.~1(b) we show the transverse velocity $V_{y}/(a\omega)$ 
versus ${\bf f}_{DC}$.
Since there is no net DC force in the transverse direction, $V_{y} = 0$ for
most values of ${\bf f}_{DC}$.
Near the steps in $V_{x}$, however, 
$V_{y}$ is nonzero, 
indicating that rectification is occurring. 
The first rectification, in the positive $y$ direction, 
occurs at the $n = 0 \rightarrow 1$ step, 
while at the $n = 1 \rightarrow 2$ step the rectification
is in the {\it negative} $y$ direction. For the higher steps,
positive rectification regions with increasing widths appear,
while near the 
$n = 7 \rightarrow 8$ step there are regions of negative rectification. 

Some of the rectification phases have very well defined heights 
of $\Delta V_{y} =  \omega a$.  These include the positive rectification 
phases at $3 \rightarrow 4$, $4 \rightarrow 5$, and $5 \rightarrow6$, 
as well as the negative rectification phases
at $6 \rightarrow 7$ and $7 \rightarrow 8$. 
On these rectification plateaus, we find that the particle
moves in only one type of orbit. 
The other rectification phases do not have a well defined
height, including $0 \rightarrow 1$, $2 \rightarrow 3$, 
as well as some portions of the $6 \rightarrow 7$ and
$7 \rightarrow 8$ regions. 
In these phases, for any fixed ${\bf f}_{DC}$
the particle jumps intermittently between different 
rectifying orbits with transverse velocities $(p/q)a\omega$.

For all of the rectifying regions,
if the polarity of the 

\begin{figure}
\center{
\epsfxsize=3.5in
\epsfbox{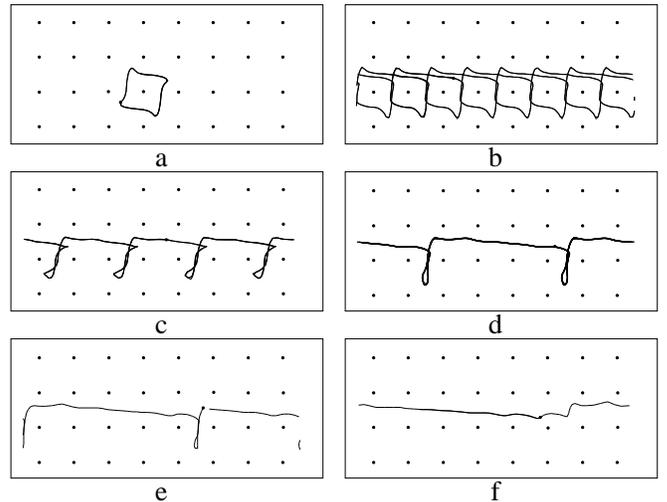}}
\caption{
Trajectories for fixed ${\bf f}_{DC}$ for different non-rectyfing regions
seen in Fig.~1. 
The black dots denote the location 
of the fixed particles or the potential maxima of the periodic substrate.
Shown are steps with:
(a) $n = 0$, (b) $n = 1$, (c) $ n = 2$, (d) $n = 4$, 
(e) $n = 5$,
and (f) $n =8$.
}
\end{figure}

\hspace{-13pt}
AC drive is reversed, $V_{x}$ remains unchanged 
while $V_{y}$ 
changes to $-V_{y}$. 
If the waiting time between DC drive
increments is increased, the results are unchanged.
The phases described here remain stable when the system is started from
a fixed ${\bf f}_{DC}$ value, and are  
not transient phenomena. We have considered various system sizes
with even and odd numbers of lattice constants $a$. 
As the system size varies, 
certain orbits 
become incommensurate with the system length
and precess spatially;
however, the velocity curves are not affected by the system size.

In Fig.~2 we illustrate representative 
non-rectifying particle orbits for $V_{x}$ steps 
of $n = 0, 1, 2, 4, 5$, and $8$. 
For $n = 0$ [Fig. 2(a)], 
$V_{x}=0$, 
and the particle 
moves in a confined square orbit which encircles one pin.
For $n = 1$ [Fig. 2(b)] there is a net motion in $V_{x}$, and the particle
circles around one pin before moving over 
to the next plaquette. For $n = 2$ [Fig. 2(c)] the orbit does not encircle a 
pin but forms a small loop 
which repeats every second plaquette. For $n = 4$ and $n=5$
[Fig. 2(d,e)], 
orbits similar to $n = 2$ occur, 
with the loop motion now repeating at every fourth or fifth plaquette,
respectively.  A similar process continues up through the $n = 7$ step. 
For the $n = 8$ step [Fig. 2(f)] and above,
the particle is moving fast enough that the transverse width
of the orbit is less than $a$, and no loops appear.

Figure 3 shows representative rectifying orbits.
The particle orbits differ below and above a given step, as
illustrated in Fig. 3(a,b) for the
$n=3 \rightarrow 4$ step.
Below the step [Fig.~3(a)], the particle moves $3a$ in the $x$ direction
and $a$ in the positive $y$ direction
in a single period.  A loop forms when the
particle moves in the $y$ direction.  Above the step
[Fig.~3(b)], the particle moves $4a$ in the $x$ direction but 
still only $a$ in the $y$ direction in one period; 

\begin{figure}
\center{
\epsfxsize=3.5in
\epsfbox{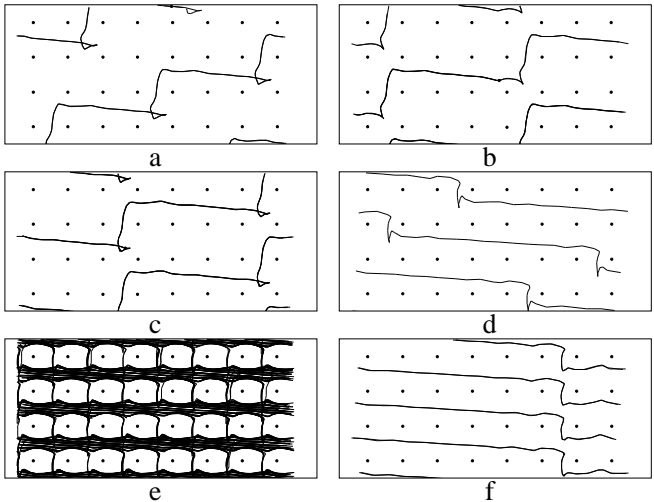}}
\caption{
Trajectories for fixed ${\bf f}_{DC}$ for different rectifying regions
seen in Fig.~1.
The black dots denote the location of the fixed particles or potential 
maxima of the periodic substrate.
(a) right before $n=3 \rightarrow 4$ step, (b) right after 
$n=3 \rightarrow 4$ step
(c) right before $n=4 \rightarrow 5$ step, (d) negative rectification phase
below the $n=6 \rightarrow 7$ 
step (${\bf f}_{DC} = 0.454$), (e) negative rectification region 
on the $n=7$ step (${\bf f}_{DC} = 0.465$), 
(f) negative rectification region on $n = 8$ step (${\bf f}_{DC} = 0.514$).    
}
\end{figure}

\hspace{-13pt}
therefore, 
$V_{y}$ does not change at the $V_{x}$ jump. 
In 
Fig.~3(c), below the $n= 4 \rightarrow 5$ step, an orbit similar to 
that of Fig.~3(a) appears, except that the particle moves $4a$ 
in the $x$ direction in each period. 
The rectifying orbit above the $4 \rightarrow 5$ step 
resembles that of Fig.~3(b), with the particle moving $5a$ in the
$x$ direction in one period.  
In Fig.~3(d) (${\bf f}_{DC} = 0.454$)  we show the negative rectification phase
below the $n = 7 \rightarrow 8$ step. Here the particle jumps $a$ 
in the negative $y$ direction every seventh plaquette
through a small kink.
We do not observe loops in the trajectories for the 
negatively rectifying phases.
Figure~3(e) (${\bf f}_{DC} = 0.465$) illustrates trajectories for a particle
in the negative rectification region close to the $n = 7 \rightarrow 8$ 
step.  The
particle does not move in a specific orbit, but jumps intermittently
over time between different orbits with $V_{y} = (p/q)a\omega$; however,
$V_{x}$ remains fixed. 
Similar intermittent trajectory patterns appear in the
rectifying phases near $n = 0 \rightarrow 1$, 
$1 \rightarrow 2$, and $2 \rightarrow 3$. 
Intermittent patterns occur only on the lower steps;
above the $n=8$ step, only stable rectifying trajectories occur, 
as illustrated in Fig.~3(f) (${\bf f}_{DC} = 0.514$). 

Rectifying phases occur for any $A$
large enough that the particle trajectory 
at ${\bf f}_{DC}=0$ encircles more than one pin.
We have measured $V_{y}/a\omega$ for the same 
system shown in Fig.~1, but with an orbit at $A = 0.42$ that
encircles four pins. 
As in Fig. 1, steps in $V_{x}$ and 
transverse rectification in $V_{y}$ appear,
but the step heights are now $\Delta V_{x} = 2a\omega$.  
We find that as $A$ is further increased, orbits
that stably encircle $p^{2}$ 
pins, 
with $p$ integer, produce
steps of height $\Delta V_{x}=pa\omega$. 
We obtain very similar results for 
substrates with a $1/r$ or $e^{-\kappa r}/r$ interaction. 

The quantization of the particle velocity on the plateaus and the
intermittent transitions between orbits 
can be explained using general properties of nonlinear maps.
Define
a map $(x,y)\rightarrow (x'+n_x a,y'+n_y a)$, from the position of the particle
at the start of a period to that at the end, where we may restrict to
$0\leq x,y,x',y'\leq a$, with $n_x,n_y$ integer.  If there is a stable
fixed point, $(x,y)=(x',y')$, then the particle translates by
$(n_x a,n_y a)$ in time $\omega^{-1}$ and so has average velocity $V_x,V_y$
quantized in multiples of $a \omega$, as found above.
If the $q-th$ power of the map has a stable fixed point, 
there are instead steps of fractional heights $(p/q) a \omega$.

As ${\bf f}_{DC}$ increases, the periodic orbit becomes unstable,
and 
a different periodic orbit with larger $V_x$ 
appears.  This new orbit will be the next stable periodic orbit
at higher drive.  The
transition to the new orbit can occur in one of three ways.
{\it (1)}  If both periodic orbits are stable simultaneously, 
the particle velocity will depend on the initial conditions
in the transition regime.   This was not observed.
{\it (2)}  The second periodic orbit 
could become stable at the same time that the first orbit becomes
unstable.  This behavior, which gives rise to infinitely sharp jumps
in $V_x$,
is not generic and hence not expected.
{\it (3)}  There can be a finite range of drive containing
no stable periodic orbits.  Over this range, the average velocity is
not quantized.  If, however, some orbits are close to stable, the 
particle will spend long times in these orbits, giving rise to intermittent
behavior.  This behavior is consistent with what we observe.

\begin{figure}
\center{
\epsfxsize=3.5in
\epsfbox{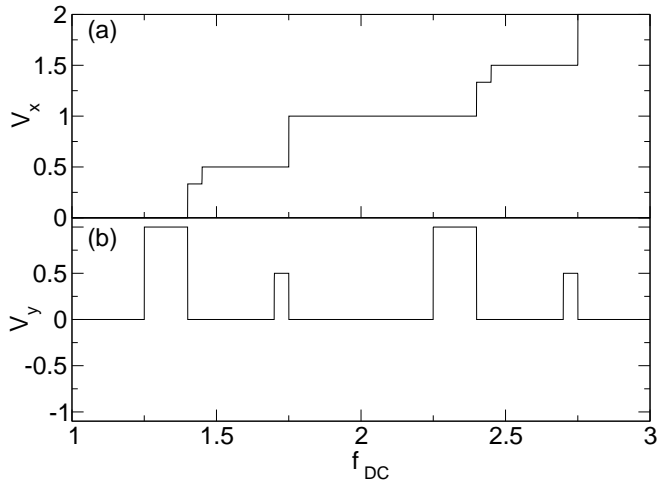}}
\caption{ 
(a) Time averaged longitudinal velocity $V_{x}$ obtained from model.
(b) Corresponding transverse velocity $V_{y}$. 
}
\label{fig:Matt}
\end{figure}

We now turn to a specific toy model illustrating some of these
ideas.  Consider a particle in a lattice of repulsive sites
with $a=1$, where the potential minima between repulsive sites are
at integer $x$ and $y$ values.  The $y$ position of the
particle is constrained 
to take only integer values, but the $x$ position can
be any real value.  To model the translation of the particle
through the lattice, 
we separate the $x$ and $y$ motion, 
so that the particle moves first ({\it i.}) right, then ({\it ii.}) down, 
then ({\it iii.}) left, then ({\it iv.}) up.  
({\it i.}): We apply the rule $x \rightarrow x+v_r$.  
({\it ii.}):  If $x$ is within $0.25$ of an integer, $x$
is set to that integer and $y$ is decremented by one.
({\it iii.}): Apply $x \rightarrow x-v_l$.
({\it iv.}): As in ({\it ii.}) except $y$ is incremented by one.
Here $v_r$ and $v_l$ are the velocity of the particle in the rightward
and leftward parts of the cycle, respectively.
In steps ({\it ii.}) and ({\it iv.}), the particle will only move
to a new $y$ position if it reaches the minima between sites
at the correct phase of the driving period, when transverse motion is
possible.  In this case, the particle slips into the next row and
the $x$ coordinate of the particle is set 
to midway between the pinning sites.  
In Fig.~\ref{fig:Matt} we show the time-averaged velocities $V_x$ and $V_y$
obtained with this model for fixed $v_l=1.15$ and increasing
$v_r$, representing increasing ${\bf f}_{DC}$.  As shown,
this simple model produces both plateaus and ratcheting behavior.  
The sharp jumps in the velocity values are
due to the discontinuity of the map function,

\hspace{-13pt}
as in case {\it (2)} above;
for smoother map functions, these 
jumps 
acquire a small but finite width.
More complicated maps 
produce
richer behavior, including occasional regimes of 
negative $V_y$.  

The ratcheting behavior in both the model and simulations
occurs near transitions in $V_x$ when the number of pinning centers the
particle passes in one period changes, making it possible for the particle
to interact asymmetrically with the pinning sites.  For a clockwise orbit,
the particle moves rapidly on the upper portion of the orbit, and is
likely to scatter off the pinning site below 
when the orbit does not quite match $na$.  On the lower
part of the orbit, however, the particle is moving more slowly, and is
likely to slip between the pinning sites above in spite of a small 
mismatch.
The particle thus tends to ratchet in the positive $y$ direction.
If the DC drive is reversed, 
downward motion should be preferred, as we observe.

Finally, we note that much of this behavior is specific to two or more 
dimensions.  Consider
 a map $x\rightarrow x'$, subject to $x+a \rightarrow x'+a$ and
${\rm d}x'/{\rm d}x\geq 0$, true for overdamped motion in one
dimension.  It can then be shown that it is not possible to have stable
periodic orbits with different values of the current.  
Systems with Shapiro steps
do not exhibit jumps.

The phases we have described should be
experimentally observable 
for vortex motion in 
superconductors with periodic pinning arrays
and Josephson-junction arrays for low flux filling
or rational filling fractions where vortex-vortex interactions are 
reduced.  
The rectifying properties discussed here
can also be a powerful tool for species separation.  For example,
colloids with different charges or DC mobilities
driven through a periodic array 
would be active ratchets since the rectified 
velocities can be controlled by  $n\omega a$, 
unlike thermal ratchets which rely on Brownian motion. 
Another system in which 
the longitudinal velocity steps
and transverse rectification could be observed is 
electrons undergoing classical cyclotron orbits in anti-dot arrays 
\cite{Weiss} 
for orbits where electrons encircle at least one anti-dot. 

In conclusion, we have found that a novel form of rectification and 
phase locking occurs for a particle with a
DC drive in the longitudinal direction 
and circular AC drive moving in a periodic substrate where the
amplitude of the AC drive is large enough that the particle can encircle 
at least one substrate potential maximum. We find that the
longitudinal velocity increases in a series of steps of height $n\omega a$.
Along these steps the particle moves in orbits that are commensurate with
the periodicity of the substrate.
For regions near the longitudinal step, transitions 
to rectification in
either transverse direction occurs. 
We have specifically demonstrated this model for vortices in superconductors
with periodic pinning arrays and overdamped charge particles. 
With a simple toy model we have shown that the qualitative features
of the phase locking and rectification can be captured. 

Acknowledgments---We thank M. Chertkov and Z. Toroczkai 
for useful discussions. 
This work was supported by US DOE W-7405-ENG-36.

\end{document}